# Think before you collect: Setting up a data collection approach for social media studies

Philipp Mayr, Katrin Weller[1]

GESIS - Leibniz Institute for the Social Sciences

Unter Sachsenhausen 6-8, 50667 Cologne, Germany

philipp.mayr@gesis.org; katrin.weller@gesis.org

## Abstract

This chapter discusses important challenges of designing the data collection setup for social media studies. It outlines how it is necessary to carefully think about which data to collect and to use, and to recognize the effects that a specific data collection approach may have on the types of analyses that can be carried out and the results that can be expected in a study. We will highlight important questions one should ask before setting up a data collection framework and relate them to the

---

different options for accessing social media data. The chapter will mainly be illustrated with examples from studying Twitter and Facebook. A case study studying political communication around the 2013 elections in Germany should serve as a practical application scenario. In this case study several social media datasets were constructed based on different collection approaches, using data from Facebook and Twitter.

# Introduction

Social media research so far is not a defined discipline. Researchers across various disciplines are interested in social media platforms and their users. Researchers with different background may focus on different research questions – and they may have their own definitions about what counts as social media research (and even about what counts as social media). To some degree this is an advantage at the current stage of studying social media, as it leaves much room for exploring approaches to address novel research questions which helps in making it an exciting topic for researchers in several fields (Kinder-Kurlanda & Weller, 2014). But this diversity also brings along a lack of standardization of approaches and thus often a lack of comparability at the current state of social media research. The present chapter will focus on the challenges that arise in designing the setup for the collection of social media data. In this context, we do not look at approaches that mainly use surveys, interviews or experiments for studying social media users and their behavior in social media environments – examples of such approaches would be Marwick and boyd (2011) who interviewed Twitter users to learn about their behavior, or Junco et al. (2011) who created an experimental setting for studying Twitter use in academic learning. In contrast to this, we focus on research that is based on datasets directly collected from social media platforms.

In general, data collected from social media could be textual content or multimedia content, user profile pages or network data, or tracked activities such as likes, shares, upvotes (see chapter 2 in this volume for an overview on data[2]). It could be data from blogs or from platforms like Facebook, Twitter, YouTube, reddit, Wikipedia, and many more. For some platforms, data can be obtained via an application programming interface (API) or via third party tools that readily provide access using the API; sometimes data has to be crawled from the website and sometimes it can be purchased through official resellers (e.g. GNIP[3] and Datasift[4]). Other researchers have come up with their own solutions to obtain social media data in a less structured format: Some researchers manually copy-and-paste selected text passages from social media platforms into excel sheets or other databases in order to create a corpus that matches their research purpose. And yet others are interested in the look of profile pages and images and may for example take screenshots to archive them as their specific data collection approach. In some cases, already existing datasets may be reused in secondary studies, although this is still rather rare – also because data sharing may be prohibited or restricted by a social media platform's terms of services (Weller & Kinder Kurlanda, 2015). Often, the chosen approach for data collection is also influenced by external factors, such as the technical limitations of a social media platform or of the data collection tool (Borra & Rieder, 2014). It might just not be possible to get the "ideal" dataset due to legal or technical restrictions and researchers do not have a choice but to work with a substitute. This does not necessarily have to be a problem and may still lead to relevant results. But researchers have to be very clear about potential limitations of their collection approach and should outline the consequences this may have for the obtained results (e.g. in terms of representativeness of their data). Knowing the boundaries of what is possible in terms of data collection is important, but it is critical not to stop thinking about its implications and to reflect on the potential biases that may arise out of it. For example, many researchers use hashtags as a convenient way to collect datasets from Twitter and this may also in some cases be the only feasible way to collect data, e.g. for an acute event. However, this may systematically exclude specific user types from the dataset, e.g. users less familiar with hashtag conversation or users who use a different set of hashtags or hashtags in different languages – or complete strains of follow-up conversations as users may no longer use the hashtag within replies to original tweets. Lorentzen and Nolin (2015) remind us in more detail of the limitations arising from hashtag based data collection approaches).

---

[2] To be added:
[3] GNIP: https://gnip.com/ (retrieved October 10, 2015)

Even when operating within some narrow limits of availability there are still choices to make: it is necessary to carefully think about which data to collect and to use; and it is important to recognize the effects that a data collection approach may have on the types of analyses that can be carried out and the results that can be expected in a study. This includes selecting the most appropriate social media channels, selecting the timeframe for data collection, constantly checking upon newly created user accounts or relevant hashtags, thinking about keywords that relate to different language communities, monitoring and documenting server outages or other technical problems. This chapter should help raise awareness of challenges around study design and data collection. For this purpose, we will highlight the most important questions one should ask before setting up a data collection framework and relate them to the different options for accessing social media data. Throughout the chapter, a specific case study will be used in order to illustrate the process of study design. The case study comes from the area of political communication in social media environments. Political communication is a frequent topic in social media research and studies that use data from Twitter for studying elections are particularly popular (Jungherr, 2016; Weller, 2014). Most of them analyze communication structures or user networks during specific cases of (national) elections (e.g. Elmer, 2013; Larsson & Moe, 2012; Towner, 2013), some also aim at predicting election outcomes (e.g. Soler et al., 2012; Tumasjan et al., 2011) – which in turn has led to some critical reflections on study design and methods (e.g. Jungherr et al., 2012; Metaxas et al., 2011) and general skepticism towards election predictions based on social media data. However, there is a potential of using social media data to monitor how people discuss political topics prior to elections and in general, how politicians interact with one another and with the public or how traditional media and social media focus on similar or different topics during elections. Research in this field includes studying politicians' interaction networks (e.g. Lietz et al., 2014), comparisons of different countries (e.g. Larsson and Moe, 2014) or close analysis of the social media campaigns of single presidential candidates (e.g. Christensen, 2013).

Even a very specific topic such as political communication during an election period can be studied in a variety of ways. Weller (2014) shows how studies on Twitter and elections vary in terms of research questions, collection period, size of the collected data set and tools for data collection. Dataset sizes can range from just single selected tweets to billions of them, from less than ten single users to networks of 200,000 user accounts (Weller, 2014).

**A practical example: social media and elections**

We will now take a closer look at the challenges for collecting data in such cases of studying political communication through social media. We use a case study which was conducted at GESIS Leibniz Institute for the Social Sciences (in cooperation with the Copenhagen Business School) and focused on political communication around the federal election that was held in September 2013 in Germany (see Kaczmirek et al., 2014 for a more detailed description of the case study). In this case study several social media datasets were constructed using data from Facebook and Twitter (a subset of the Twitter dataset has also been archived for reuse, see Kaczmirek and Mayr, 2015).

The project goal was to examine various aspects of communication structures in online media and to investigate how such data can add new insights in comparison to existing data from surveys and (traditional) media analyses (Kaczmirek et al., 2014). The project was tied to the broader framework of the German Longitudinal Election Study (GLES[5]), a long term research project that examines the German federal elections in 2009, 2013, and 2017 with the aim of tracking the German electoral process over an extended period of time (Schmitt-Beck, Rattinger, Roßteutscher & Weßels, 2010). Data used in the GLES project includes surveys, media content analyses, and interviews with election candidates. The overall aim was to supplement the GLES candidate study – which is based on interviews – with new information about the candidates retrieved from social media sources. Another idea was to complement the traditional media corpus analysis of GLES (were different traditional mass media channels are analyzed) with an analysis of important topics as discussed in social media. As we will see below, we decided to do this based on data from Twitter and Facebook. We will use this exemplary case to illustrate some more general strategies for social media studies.

---

[5] http://www.gesis.org/en/elections-home/gles/

# Strategies for data collection

The first steps in setting up a social media study will usually be to formulate a research question and to then decide upon the most suitable data that will allow answering this question. It has been criticized that a lot of 'big data' studies are data driven, i.e. starting with a given dataset rather than with a research question or theory – and critical reflections are emerging on how such data-driven approaches affect knowledge production (e.g. Schroeder, 2014).

Starting with a given dataset and building the research questions around it, can make a lot of sense in some cases. This exploratory design may be useful for mapping out the properties of a specific social media platform and is thus applied in cases where one first needs to understand the overall usage scenario or the user behavior within a specific platform (as done by e.g. Cha et al., 2007 for YouTube; Weninger et al., 2013 and Singer et al., 2014 for Reddit). But in most cases, it is indeed recommended to start with the specific research question and then to think about the ideal dataset that would be needed to answer it. In the next steps, one may have to lower the expectations: the ideal dataset may not be possible due to, for example, technical, legal, or ethical limitations. For example, the ideal dataset for some research question might be all tweets ever sent on Twitter from locations in Germany. Unfortunately, it is not possible to collect tweet searches retrospectively via the public Twitter APIs[6], and even if one can afford buying such a large dataset from the official Twitter data reseller GNIP there still is the fact that only very few tweets are geo-coded, so that it is not easily feasible to identify tweets sent from Germany. In such cases, one has to find a way to approach the best possible dataset and acknowledge some drawbacks and limitations. Over time, the research community is learning which kind of data can be crawled from specific social media platforms, and which not, and is exchanging best practices and lessons learned (though it has to be kept in mind that as social media platforms and their APIs may change, all this expertise has to constantly evolve, too). Still, it is important to always envision the ideal dataset and then reduce it to the best one given the current limitations. If researchers simply work with the same kind of data which has been used before and proved to be easily accessible, there is a risk that they miss opportunities for creating better data collection approaches. For example, working with a Twitter dataset collected for a specific hashtag has become common practice, so that some researchers might forget to think about whether different synonymous keywords would have been more appropriate entry points for data collection.

---

[6] API is short for Application Programming Interface. For Twitter's APIs see Twitter's website on technical information (Twitter, no date a and b) and Gaffney & Puschmann (2014).

Thinking about the ideal dataset should of course also include asking whether social media will really provide the best possible data source – or whether other data (e.g. experiments, survey data, content from traditional mass media) would be more appropriate. In the following we will introduce a set of questions that are critical to any data collection approach in social media research. The initial question should be:

1. Which social media platforms would be the most relevant for my research question? (Single platform vs. multi-platform approach)

When this is decided, the next step will be to prepare data collection from the selected platform(s), while asking the following questions:

2. What are my main criteria for selecting data from this platform? (Basic approaches for collecting data from social media)

3. How much data do I need? (Big vs. small data)

4. What is (unproportionally) excluded if I collect data this way? (Collection bias)

We will now take a closer look at these questions and the possible strategies for data collection related to them.

**Single platform and multi-platform studies**

Many current social media studies focus on a single social media platform, with Twitter and Facebook being most prominent (Weller, 2015). For research that aims at gaining a deep understanding of a specific platform, this single platform approach is self-evident and appropriate: in order to, for example, fully understand how Twitter users make use of retweets (e.g. boyd et al., 2010) or hashtags it is most crucial to collect data from Twitter. But even in these cases, a comparison with other platforms would be desirable in order to proof whether the observed phenomena are unique to Twitter or in line with results from other contexts. Quan-Haase and Young (2010) demonstrate the value of comparisons across platforms in social media research.

While some research focuses on understanding a specific platform, other studies look into selected phenomena such as political activism (e.g. Faris, 2013; Thorson et al., 2013), disaster response (e.g. Bruns & Burgess, 2014; Vieweg et al., 2010), scholarly communication (Haustein et al., 2014) or journalism (e.g., Papacharissi, 2009). Often these cases are narrowed down to how a specific platform was used in a specific situation, like Twitter during the London Riots, Facebook during the presidential election, Flickr for interacting with street art, and YouTube for e-learning. All these examples would promise interesting insights. But in the long run, we also need more approaches that consider the role of different platforms within the broader landscape of traditional and new media formats, i.e. how different social media platforms are either interrelated or complement each other – as illustrated by Quan-Haase and Young who also argue for different needs being met by different platforms (Quan-Haase and Young, 2010). A lot of topics may not be discussed in isolation on just one platform. URLs may be included to explicitly link between different platforms: tweets may include links to Facebook, Facebook posts may reference YouTube videos, Wikipedia articles may reference blog posts etc. Memes (Zappavigna, 2012) may spread from one social media platform to the other. For many topics, the full picture will only become visible by including data from more than one social media platform. On the other hand, social media users may purposefully choose one platform over the other for different needs (Quan-Haase & Young, 2010). This means that different platforms may be used for different kinds of communication, and that some platforms may be more suitable for studying specific topics than others.

In our use case (studying online communication during the German federal election 2013) we also had to think about which social media platforms we wanted to study. We started by considering the social media platforms which are the most popular in Germany. As we wanted to collect data about election candidates, we focused on the platforms that were most broadly used by this group of people: Facebook and Twitter. For the purpose of collecting data about politicians' communication patterns we thus planned to include both of these platforms. Because of its greater ability to connect different forms of publics (Schmidt, 2014) and because of the feasibility to discuss topics spontaneously based on hashtags, Twitter was selected as a suitable platform to look for discussions around the electoral campaigns (which might be compared to contents of mass media coverage). After this was decided, we had to move on to clarify the exact setup for data collection.

**Basic approaches to collect data from social media**

There are a number of ways that data can be composed and collected. First of all, for every study one has to decide upon the timeframe for collecting data. The selected timeframe may heavily influence the results, as for example demonstrated by Jungherr et al. (2012) for the case of election prediction, where different data collection periods lead to different predictions about election outcomes. Time is a fundamental dimension that needs to be considered in all data collection approaches, e.g. should data be collected for single hours or maybe for months or even years? The timeframe then needs to be considered in combination with the basic strategies that can underlie data collection setups. The most common criteria for data collection are:

a. *Based on user accounts*. Given the case that we know a complete group of users, it might be desirable to collect data for all persons or instances in that group. This could be all soccer clubs within a country (Bruns et al., 2014), all Dow 30 companies, or – as in our example – all candidates running for a specific election. It is extremely helpful if a list of all individuals belonging to a group exists already, as this is the closest you can get to a full sample in social media research. If you can identify for example all members of parliament who are on Twitter, you have the ideal starting point for comparing them. However, identifying all members of a group is not always trivial, and in some cases – as we will see below – the outlines of the group may be fuzzy and decisions will have to be made about who to include or not. In many cases, it will not be possible to identify everyone belonging to a specific group, for example all people in a country who are eligible to vote. Cases in which we can assemble a full sample will thus most likely refer to some sort of elite users, rather than broad groups of people.

In our case study, it was possible to identify more than 2,000 candidates running for the German elections and to check if they had a Twitter or Facebook profile (more details below).

b. *Based on topics and keywords*. A very frequent approach is to collect social media content based on topics, e.g. for a specific event (like elections or sports events) or a general topics that are being discussed by a group of people (like same sex marriage). Especially on Twitter, topical discussions are often labeled with specific hashtags, but other platforms also enable the users to apply content-descriptive metadata like tags or keywords. These may be used as a criterion for searching and collecting social media data. In other cases, the full texts of social media contents (tweets, Facebook posts, blog posts, comments etc.) can be used for collecting all cases that include a specific word. However, in many cases it is difficult to achieve 'completeness' in data collection when using text-based collection approaches. People may use different vocabulary to refer to the same topic, or the topic may not clearly be mentioned in very short posts at all. For example, on Twitter, some people may use one or more designated hashtags when commenting on a current event (e.g. #WorldCup2014 for the FIFA World Cup in 2014), others may use different hashtags (e.g. #Brazil2014, or also for example hashtags in different languages), or some may mention the event without using any hashtag and some may comment on the event even without saying its name.

When setting up a data collection approach based on keywords or other full text searches, it is important to document the choice of search terms and to consider potential alternatives. In some cases, it may be possible to collect entire threads of discussions even if only one single comment included a word that matched a query, which can lead to a more complete data collection approach.

In our case study, data was collected by utilizing a series of keywords (in addition to the collection approach based on users). We will describe this in more detail below.

c. *Based on metadata*. In some cases, data is collected based on some other structural criteria, which we call metadata in this context. This should reflect anything that is neither based on a person's or account's name nor on any content features based on semantics (keywords, hashtags). Examples for metadata that can be used for data collection include, but are not limited to, geo-locations (e.g. all status updates published in a specific country), timeframes (e.g. all status updates posted on a Sunday), language (e.g. all status updates in languages other than English), or format (e.g. only retweets, only status updates that include a URL or an image, all YouTube vides longer than 3 hours). Their availability depends on the selected social media platform and their data access policies.

d. *Random sample*. Finally, it may be possible and useful to collect a random sample of data from social media platforms. This is particularly useful for studies that want to investigate general characteristics of a social media platform (and not focus on a specific topic or user group). Some APIs may directly offer access to a random set of contents.

When collecting data based on one of the previous approaches, the resulting dataset may also be too large for some types of analysis (e.g. based on software limitations) or for some data infrastructures and may thus require some post-collection sampling.

**Big data or small data?**

Big data has become a buzzword in different contexts and is also often used to refer to social media studies. Indeed, with the growing number of social media users, the rate at which content is being shared also increases. There are several examples of studies that have collected data from large numbers of social media users, e.g. Kwak et al. (2010) and their network of more than 40 million Twitter users. And yet there is no shared definition about what counts as "big" in a social media research context (see Schroeder, 2014 for an approach). People may probably quite easily agree that a given dataset of for example 10 user profile pages constitutes an example for "small" data, but if these are heavy users of a certain platform who accumulate millions of status updates the perspective may change (see chapter on thick data, this volume). It is certainly more common to refer to the number of units for analysis (user accounts, nodes in a network, content units such as tweets or Facebook posts, actions such as likes or views) than to the size of the storage needed for handling the data (e.g. in gigabyte or terabyte). Still, questions of data storage and processing infrastructure have to be carefully considered when dealing with social media data.

There now are a couple of critical reflections on big data research and its drawbacks, focusing for example on representativeness, ethical issues and the role of APIs as black boxes. boyd and Crawford (2012) collected "six provocations" for big data researchers to remind them of research ethics as well as the potential lack of objectivity. Ethical challenges of working with user data without explicit consent and with limited possibilities for anonymization are being discussed for specific case studies, e.g. by Zimmer (2010). Bruns (2013) adds arguments about the lack of documentation of collection methods resulting in lack of replicability of many studies. Tinati et al. (2014) discuss the changing nature of social media platforms and the effects this has on data collection and analysis. Lazer et al. (2014) demonstrate how other changes, namely in user behavior, can also lead to problems with big data analyses. All this has practical implications for the data collection setup. And as little general guidelines exist, it is upon the individual researcher to figure out for him/herself how much data will be needed for answering a specific research question.

In addition to big data and small data, several other phrases have been used to refer to social media data and to highlight the specific qualities instead of the quantity, e.g. "compromised data" (Langlois et al., 2015). In many cases, the essential question is not about the actual size of a dataset – in the end it comes back to how the dataset has been composed, or what criteria were applied in order to collect it.

**Dealing with collection biases**

Many approaches to data collection induce a specific bias to the dataset (see Chapter 52 for a more detailed discussion of biases[7]). Ruth and Pfeffer (2014) also discuss a variety of sources for bias in social media research, and Bruns and Stieglitz (2014) do so for the case of Twitter in particular. Some common sources for biases are:

a. *Biased social media populations.* In many cases little is known about the exact population of a social media platform, e.g. in terms of gender, age, location, or other factors such as political orientation, education etc. In most cases we can assume that social media platforms are not representative of a general population (e.g. of a specific country). Unless the relation is known, it is rarely possible to make statements beyond the platform users. Also, different social media platforms address different user populations and may not easily be compared.

b. *Access restrictions:* Most platform providers somehow restrict the access to their users' data. Often these restrictions are not completely transparent. For example, Morstatter et al. (2014) question whether the data provided through the Twitter API are representative of Twitter in total.

c. *Sampling biases.* The different approaches to data collection described above may also induce certain biases. For example, collecting tweets based on geo-codes only includes tweets by users who have deliberately chosen to share their geo-location, a sub-group which may not be representative of all Twitter users.

---

[7] To be added

# Case study for data collection[8]

In the following we will outline a case study which has been undertaken in 2013. More details of this study can be found in the working paper Kaczmirek et al. (2014).

We have briefly provided single examples drawn from the case study in the previous sections. Now we will give a more comprehensive account about how the working group approached data collection to illustrate some of the practical challenges we encountered – especially highlighting those challenges occurring before the actual data collection would begin.

The goal of Kaczmirek et al. (2014) was to collect social media communication which is closely related to the last German Bundestag elections on September 22nd, 2013. The corpus should enable the team to study both the election candidates and their behavior in social media environments (in contrast to other media channels) and different topics that were debated by social media users during the pre-election period. "To this end we constructed different data sets which we refer to as the 'Facebook corpus of candidates' (a corpus which shows how politicians communicate and represent on Facebook), the 'Twitter corpus of candidates' (a corpus which shows how politicians communicate and represent on Twitter), the 'Twitter corpus of media agents' (a corpus which shows how media agents and journalists communicate and represent on Twitter), the 'Twitter hashtag corpus of basic political topics', the 'Twitter hashtag corpus of media topics', and the 'Twitter hashtag corpus of the Snowden affair'. The first corpus includes data collected from the Facebook walls of candidates for the German Bundestag. For the other corpora we collected Twitter data. The last corpora contain tweets identified by a list of hashtags which was constructed following a topical approach" (Kaczmirek et al. 2014, p. 9). This topical approach was intended to compare different media channels e.g. with the study of political topics in classical media in GLES. "Technically, we collected tweets sent from account names of our lists (see below), tweets in which those names

---
[8] The section "Case study for data collection" is a slightly extended version of sections in a previous working paper (see Kaczmirek et al. 2014. In this section we will qoute major parts directly from the working paper.

were mentioned (i.e., which included the @-prefix) and tweets which matched our hashtag lists (i.e., which included the #-prefix)." (Kaczmirek et al. 2014, p. 9)

**First preparations for data collection: Setting up a list of candidates for the German Bundestag**

For the goal of studying social media communication by election candidates, Kaczmirek et al. (2014) had to start by setting up the list of relevant persons and their social media accounts. This means that in this case, it was suitable to work with a person-based approach for data collection. Principally, it would have been possible to use this approach for all candidates running for the 2013 election. However, some additional manual effort was needed to set up the list of candidates, as by the time the data collection from Twitter had to begin in real time the official lists of candidates had not been published yet. "Although an official list of Bundestag candidates is published by the Bundeswahlleiter (federal returning officer) six weeks before the elections, we decided to investigate the candidate names ourselves. We did this in order to be able to start data collection of social media data simultaneously to the start of the GLES media content analysis in June 2013 and in order to collect data sufficiently in advance before the election would take place" (Kaczmirek et al. 2014, p. 9) (this means that in this case the decision about how long the data collection period should be was based on the desire to be able to match the data with another available dataset in a given collection period).

The working group thus wanted to construct a list of names of the relevant candidates which could be used as the starting point for the search of the social media accounts for both candidate corpora. "Relevance was defined as the reasonable likelihood of becoming a member of the Bundestag. We refer to this list as the list of candidates although the complete number of overall candidates was higher. The data was collected in a two-stage process.

In the first stage, the names of the Bundestag candidates and details of their candidature (list or direct candidature; constituency) were searched on the webpages of the party state associations (six parties x 16 state associations). If the candidates were not announced online, the names were requested via email or telephone call at their press and campaign offices. Since the direct candidates are elected separately in every constituency and since the party congresses, where the list candidates are elected take place at different times, our list of candidate names was continuously extended.

In the second stage, the Facebook and Twitter accounts of the candidates were identified based on the list of candidates. In addition to the internal Facebook and Twitter search function, the list of social media accounts of current members of parliament on the website pluragraph.de was useful. Furthermore, several of the politicians' or parties' websites linked to their social media accounts.

We applied the following criteria to verify that the accounts were related to the target person: (1) Is a reference to the party, for example a party logo visible? Are Facebook friends and Twitter followers members of this party? (2) Do the candidate's personal or party website link to the profile? (3) Can the candidate be recognized via image or constituency (for direct candidates)? Where available, the verified badge in Twitter was used to select the correct account of a candidate in cases of multiple available accounts.

If the candidate had an account which he or she used for private purposes in addition to his professional account[9], only the professional account was included in our list. During our search for the accounts, this problem occurred primarily with Facebook accounts. Since a list of candidates of the 2009 Bundestag election was already available from the 2009 GLES candidate study, we also searched Facebook accounts for these candidates." (Kaczmirek et al. 2014, p. 9-10)

In the end the working group identified a list of persons who would run for the election (n=2,346). On Facebook the working group was able to collect information from 1,408 Facebook walls. On Twitter the working group followed a set of 1,009 candidates (and added 76 other agents, for example, journalists, for our additional research goals).

---

[9] We could only identify accounts that were publicly available. We did not search for accounts for which the account holder had decided to make it a "private" account in the sense that it is not shared with the public.

Kaczmirek et al. (2014) have used an approach based on a list of user accounts as described of one of the possible options for data collection outline above. So far the working group has seen that even for a defined group of persons realizing this approach may require considerable effort and has to be done manually. The main challenge is in identifying the actual accounts and verifying that they are correct and official. Sharing archived lists of identified user accounts (as done by Mayr and Kaczmirek, 2015 and recently for another case by Stier 2016) thus is of value for other researchers who might be interested in the same set of accounts and reduces manual effort.

In a next step, we describe which other approaches in addition to the list of candidates was used for data collection in our context.

### Defining different entities as lists: e.g. gatekeepers, information hubs and hashtags

"Since Twitter is a fast medium which takes up and redistributes new information quickly, it is likely that conventional media also use Twitter as a data source. We assume that conventional media select information from Twitter and refine and redistribute the topics over the more conventional media" (Kaczmirek et al. 2014, p. 10). The 'Twitter corpus of media agents' was intended to reflect this. "We refer to the individuals who would follow such an information gathering approach as 'gatekeepers' and searched for them among journalists and editors.

In a first step, we identified journalists and editors working in internal political divisions of national daily newspapers and magazines and searched their Twitter accounts. The leading principle in selecting the media sources was whether they were included in the print media content analysis of GLES. The result of this first step is a list of all Twitter gatekeepers of conventional media.

In a second step, we retrieved all accounts that the gatekeepers followed. The assumption behind this approach is that the gatekeepers themselves track what we call 'information authorities'. The information authorities push topics into Twitter and it is likely that they play a central role in shaping the agenda on Twitter. In order to be counted in the list of information authorities we introduced the criterion that at least 25 percent of the gatekeepers have to follow the account. The list is extended by accounts which are followed by at least 25 percent of the journalists or 25 percent of the editors.

These data may prove useful to supplement research related to both the social media content analysis (…). Furthermore, the communication, bonds and agenda-setting among gatekeepers and information authorities themselves can be the target of research. The gatekeepers and information authorities constitute the source (…) for the Twitter corpus of media agents." (Kaczmirek et al. 2014, p. 10)

"In defining (…) the Twitter hashtag corpora, we took an alternative approach which was not restricted to communication around specific Bundestag candidates or journalists. To gain information about the political communication of the population on Twitter, we used thematic hashtags. Here, we defined three procedures which serve to generate three lists of relevant hashtags." (Kaczmirek et al. 2014, p. 11) The working group divided the hashtag corpora into "basic political topics and keywords", "media content" and a case study "NSA / Snowden".

The list "basic political topics and keywords" "is comprised of the common hashtags (abbreviations) of parties in the Bundestag (…) or of parties which are known to communicate substantially via social media (e.g., the party "Piraten"). The list is complemented with the names of the party top candidates as hashtags (e.g., #merkel). A collection of hashtags for the parliamentary elections in general (e.g., #wahl2013 [#election2013]) completes the list. These hashtags comprise different conjunctions and abbreviations of election, Bundestag, and the year 2013" (Kaczmirek et al., 2014, p. 11 and appendix).

The list 'media content' "is based on the coding scheme of the media content analysis of GLES (GLES 2009). Wherever reasonable, one or more hashtags were generated for each code in the coding scheme (e.g., the coding scheme used 'Landtagswahl' and the corresponding examples for the hashtags included #landtagswahl, #landtagswahl2013, #landtagswahl13, #ltw). The main challenge in setting up this list was that not all issues could be transformed into meaningful hashtags because topics would become too broad and produce more noise in the data than valuable content. This list is therefore subject to a higher selectivity and less objective than the first list." (Kaczmirek et al., 2014, p. 11)

## The lack of flexibility in the fixed list approach

"With the election the party AfD (Alternative for Germany) made an important leap forward. In the initial concept we had not foreseen these events. Therefore, communication about and from AfD candidates is not initially included" in the candidate corpus from Twitter "but 15 AfD candidates were added on the 27th of November 2013 to the Twitter data gathering procedure. While it is possible to collect tweets from these accounts back to the start of our data collection efforts, this is not possible for @-messages to these users or tweets including their names as a hashtag. Unfortunately, we are unable to add the Twitter communication for the other corpora because monitoring could only be implemented in real-time making it impossible to capture past events." (Kaczmirek et al., 2014, p. 12) The only option to include the missing data would be to buy them from official resellers.

"Because Facebook posts are more persistent we were able to include data of the candidates of the party AfD. The Facebook walls of AfD candidates (…) were re-fetched and are part of the corpus definition." (Kaczmirek et al., 2014, p. 12)

## Reusing lists to automatically crawl data

### Collecting data from Facebook

For the first candidate corpus, "the Facebook data were collected and analyzed using the purpose-built software application Social Data Analytics Tool" (SODATO[10], see Figure 1 below, Hussain & Vatrapu, 2014). "This tool allows examining public interactions on the Facebook walls of Bundestag

---

[10] http://cssl.cbs.dk/software/sodato/

candidates by extracting several conceptual core types of information: Breadth of engagement (on how many Facebook walls do individuals participate); depth of engagement (how frequently do individuals participate); specific analytical issues such as modes of address (measured use of first person, second person, and third person pronouns); the expression of emotion (positive, negative, and neutral sentiment); the use of resources such as webpages and YouTube videos; verbosity; and extent of participation. In the case of modes of address and expression of emotion, one can also examine how they evolve over time." (Kaczmirek et al., 2014, p. 13)

"To fetch the relevant social graph and social text data from the Facebook walls, we used SODATO. SODATO uses and relies on Facebook's open source API named Graph API. SODATO is a combination of web as well as Windows based console applications that run in batches to fetch social data and prepare social data for analysis. The web part of the tool is developed using HTML, JavaScript, Microsoft ASP.NET and C#. Console applications are developed using C#. Microsoft SQL Server is used for data storage and data pre-processing for social graph analytics and social text analytics. A schematic of the technical architecture of SODATO is presented in Figure 1." (Kaczmirek et al., 2014, p. 14)

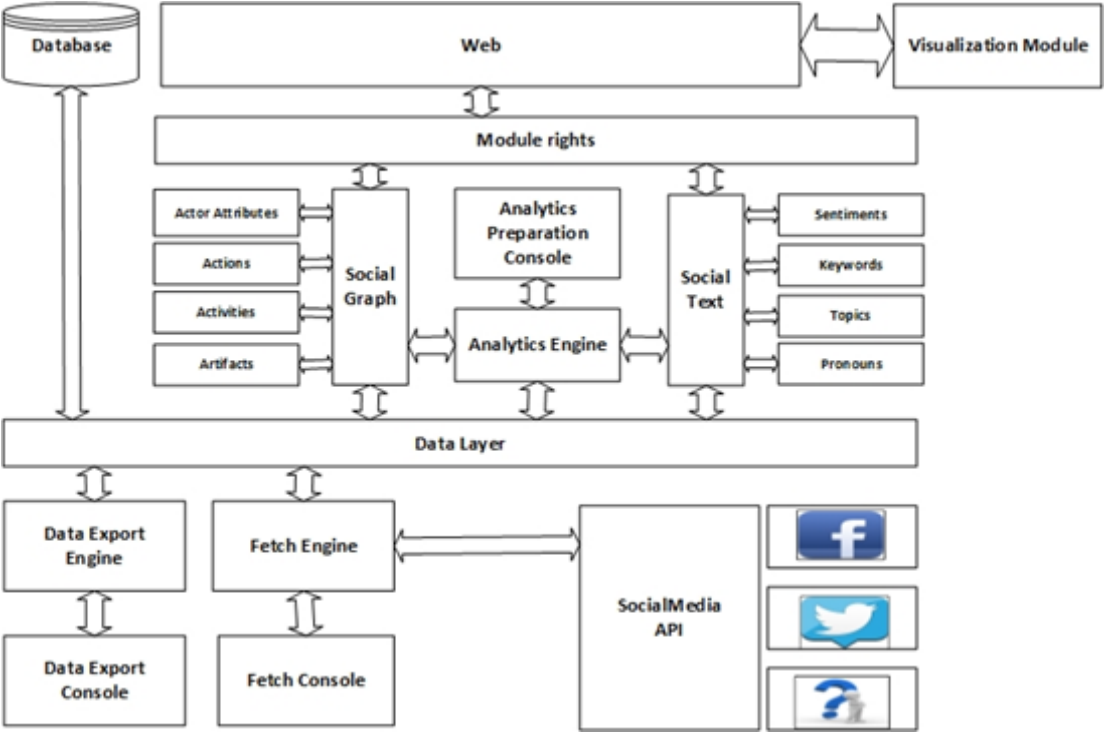

Fig. 1: Schematic of the technical architecture of SODATO (Kaczmirek et al, 2014)

*Collecting data from Twitter*

"In the following we describe the technical aspects of creating the Twitter corpora. The Twitter monitoring builds upon previous work by Thamm & Bleier (2013). As outlined above Twitter data is used to build different corpora. (…) Applying the list of candidate names which have an active professional Twitter account in the 2013 elections we used the Twitter streaming API[11] to receive messages directly from these candidates as well as the retweets of and replies to their messages. (…) For that purpose we developed a software component called TweetObserver that is instantly reading the stream from Twitter resulting from our query in a stable manner" (see Figure 2). "The software needs to register as a Twitter application in order to continuously receive update events for the requested items from the Twitter service. For each account the search query includes the account ID and the name, so that the application is geared towards receiving tweets from a certain account as well as any mentioning of its name. The software was implemented in Java and relied on the Twitter library twitter4j[12]. The software is connected to a MongoDB in which we store the data in JSON format. In the following we describe the data structure of the tweets in the Twitter data set." (Kaczmirek et al., 2014, p. 17)

As it is unclear if the TweetObserver software is always able to receive all tweets from the requested accounts the working group introduced a simple quality proofing mechanism. To assess the completeness another component called ObserverTester was introduced that controls the TO by automatically creating tweets at defined intervals matching its search criteria. Since all generated tweets need to be stored by the first program, the completeness is estimated as the difference between the created and the stored tweets (see Fig. 2). $TO_{1...n}$ are instances of the program that observes different twitter accounts.

---

[11] https://dev.twitter.com/streaming/overview
[12] http://twitter4j.org

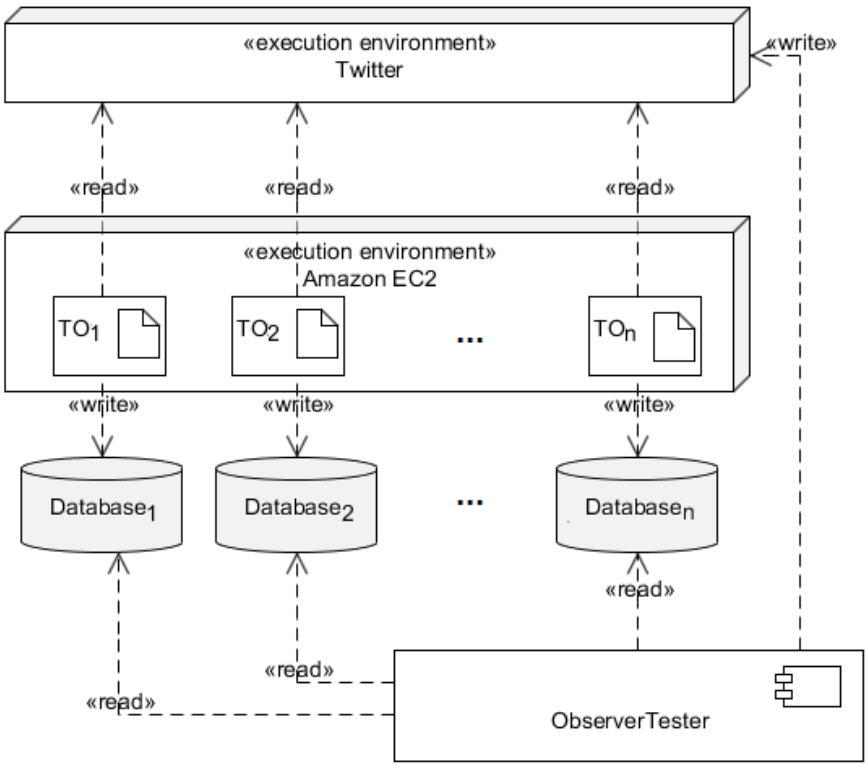

Fig. 2: Schematic of the technical architecture of a Twitter TweetObserver

In table 1 the data structure of the tweets in the Twitter data set is explained. The collected tweets are in JSON format and contain at least the attributes presented in table 1.

Table 1: Selected attributes of tweets available in JSON format (adapted from Kaczmirek et al, 2014).

| Attribute | Description | Example |
|---|---|---|
| _id | tweet ID | 446226137539444736 |
| userid | numeric user ID | 630340041 |
| screenName | alpha numeric user ID | lkaczmirek |
| createdAt | date of tweet | 2014-03-19T11:08:00Z |
| tweettext | text of this tweet | @gesis_org is offering #CSES data, providing electoral data from around the world: https://t.co/phtZgGcljs |
| hashtags<br>• start<br><br><br><br>• end<br><br>• text | internal collection of hashtags with the following attributes:<br>• index of the start-character (the position in the tweet text as a number, the first letter equals index zero)<br>• index of the end-character (the position in the string as a number)<br>• the tag itself | <br>• 23<br><br><br><br>• 28<br><br>• cses |
| mentions<br>• start<br><br><br>• end<br><br><br>• id<br>• screenName<br><br>• name | internal collection of user mentions with the following attributes:<br>• index of the start-character (the position in the string as a number)<br>• index of the end-character (the position in the string as a number)<br>• user ID of the mentioned user<br>• screen name of the mentioned user (account name)<br>• name of the mentioned user | <br>• 0<br><br><br>• 10<br><br><br>• 145554242<br>• gesis_org<br><br>• GESIS |

# Conclusions and Outlook

A lot of the decisions that need to be made when setting up the data collection for a social media study rely on the considerations of the individual researcher and his/her team. So far, there are often no or very few guidelines that can help in this process. Social media research is still on its way towards establishing methodological standards or best practices.

In the exemplary case study we have seen that before the automatic crawling and collecting of social media data can begin, a lot of underlying research is necessary. Before data can be collected, different preparations may be necessary, such as strategic decisions about the period of data collection and the search criteria for collecting data. We have shown how this can be approached this for different types of datasets, a data collection approach based on lists of user accounts or based on topics and corresponding hashtags. The different types of collected data sets allow for dealing with different research questions.

Due to restrictions in the Twitter API it is not possible to collect some types of data retrospectively. In the presented case study this meant, that in one case it was not possible to fully react to some unforeseen event (the unexpected growth of a new political party in Germany, which was not anticipated when setting up the data collection approach). Other projects will have to face different challenges based on technical restrictions.

A dimension we have only touched upon very briefly in this chapter, but which also plays a huge role in practice, are the legal and ethical challenges for working with social media datasets (both are increasingly being discussed in the research community). In this presented case, legal restrictions and ethical considerations mainly played their most crucial role after data collection, namely when it came to approaches for sharing the collected datasets. We wanted to make as much as possible of our datasets available for reuse.

In the end, the following data was shared (see the dataset published as Kaczmirek & Mayr, 2015): (1) A list of all candidates that were considered in the project, their key attributes and if available the identification of their Twitter and Facebook accounts. (2) A list of Tweet-IDs which can be used to retrieve the original tweets of the candidates which they posted between June and December 2013. It includes the Tweet-ID and an ID identifying the candidate. According to the Twitter terms of services[13] it was not possible to publish the full Twitter data (tweets plus metadata in its original format). During discussions at the GESIS data archive it was furthermore decided that the Twitter data may contain potentially sensitive information such as political opinion and maybe even information on voting behavior. It was decided to limit the shared dataset to data from actual election candidates, and for privacy reasons tweets from the general Twitter population are currently excluded.

Even publishing a small subset of a collected social media dataset still is an achievement; in most cases social media datasets are currently not being shared at all. Together with a white paper about the underlying data collection approach (Kaczmirek et al., 2014) a shared dataset constitutes a first step towards more detailed documentation for social media research projects. Both, documentation and data archiving, certainly need to be extended for social media research in general in the future, in order to make decisions behind data collection understandable and data collection approaches reproducible.

## Acknowledgements

We thank our colleagues who were part of the working group for the project "PEP-TF: Social Media Monitoring of the Campaigns for the 2013 German Bundestag Elections on Facebook and Twitter", which was the basis for the case study presented in this chapter. The project was initiated at GESIS – Leibniz Institute for the Social Sciences. The project goals and the conceptualization of the data were developed by GESIS, who also undertook data collection on Twitter. The project partner at the Copenhagen Business School, Computational Social Science Laboratory (CSSL), Department of IT Management used their Social Data Analytics Tool (SODATO) to conduct data collection on Facebook. The project was led by Lars Kaczmirek (for social science dimensions) and Philipp Mayr (for computer science dimensions). Thanks to Manuela Blumenberg and Tobias Gummer, who developed the project goals under the supervision of Lars Kaczmirek. Alexander Wenz helped in researching the politicians and their accounts. Together, they constructed the source information for data collection. Arnim Bleier and Mark Thamm contributed the technical realization of collecting Twitter data with TweetObserver. Kaveh Manshaei helped with resolving the shortened URL in the Twitter dataset into the original URL. Ravi Vatrapu was the supervisor in Copenhagen. Abid Hussein conducted the data collection on Facebook and developed the necessary tools under the supervision of Ravi Vatrapu. The software SODATO was developed by our project partners at the Copenhagen Business School, Computational Social Science Laboratory (CSSL), Department of IT Management. We thank Ravi Vatrapu and Abid Hussein of CSSL for their collaboration and support. Katrin Weller, Katharina Kinder-Kurlanda and Wolfgang Zenk-Möltgen developed that framework for archiving first subsets of the collected Twitter data, Thomas Ebel and colleagues at the GESIS data archive furthermore supported the practical archiving process. Finally we thank Christof Wolf as the scientific advisor of the project. The project started in March 2013.

---

[13] https://twitter.com/tos